\documentclass{jfm}
\usepackage{epstopdf, epsfig}
\usepackage{graphicx}
\usepackage{xcolor}
\usepackage{ulem}

\shorttitle{Critical shear rate and torque stability condition for a particle  in a fluid flow}
\shortauthor{A. Kudrolli, D. Scheff and B. Allen}

\title{Critical shear rate and torque stability condition for a particle resting on a surface in a fluid flow}

\author{Arshad Kudrolli, David Scheff, and Benjamin Allen}
\affiliation{Department of Physics, Clark University, Worcester, MA 01610
}%

\begin{document}

\maketitle
\begin{abstract}
We advance a quantitative description of the critical shear rate $\dot{\gamma_c}$ needed to dislodge a spherical particle resting on a surface with a model asperity in laminar and turbulent fluid flows. We have built a cone-plane experimental apparatus which enables measurement of  $\dot{\gamma_c}$ over a wide range of particle Reynolds number $Re_p$ from $10^{-3}$ to $1.5 \times 10^3$. The condition to dislodge the particle is found to be consistent with the torque balance condition { after including} the torque component due to drag about the particle center. The data for $Re_p < 0.5$ is in good agreement with analytical calculations of the drag and lift coefficients in the $Re_p \rightarrow 0$ limit. For higher $Re_p$, where analytical results are unavailable, the hydrodynamic coefficients are found to approach a constant for $Re_p > 1000$. We show that a linear combination of the hydrodynamic coefficients found in the viscous and inertial limits can describe the observed $\dot{\gamma_c}$ as a function of the particle and fluid properties.
\end{abstract}

\begin{keywords}
Erosion; Granular; Onset; Shear flow
\end{keywords}

\section{Introduction} 
The threshold condition needed to dislodge particles, which are initially at rest on a surface,  due to a fluid flow is important in a wide range of physical systems and industries. Examples include wind blown dynamics of sand dunes, erosion of sediments and rocks on river beds and ocean floors, deposition of proppants in hydraulic fracturing of shales, and drug delivery via inhalation. In spite of a long standing interest in such problems (see Shields 1936; Buffington \& Montgomery 1997), the conditions under which particles are dislodged by a fluid flow are not well established quantitatively.  The Shields number, given by the ratio of the hydrodynamic drag and gravitational force acting on the particles at the surface, is often used to characterize the physical conditions at the threshold of motion~(see Shields 1936; Wiberg \& Smith 1987; Buffington \& Montgomery 1997). When this number exceeds a value corresponding to an effective friction, the fluid is considered to dislodge the particles. { This appears to imply that the condition to dislodge a particle can be characterized by considering the mean forces acting on the particle alone~(see Phillips 1980; Wiberg \& Smith 1987).} Accordingly, the recorded Shields number at the threshold of motion has been reported over a wide range of Reynolds numbers estimated at the particle scale and shows broad trends with considerable scatter~(see Buffington \& Montgomery 1997) that depend on bed preparation, particle size and degree of exposure to the fluid flow~(see Charru, Mouilleron \& Eiff 2004; Charru, {\it et al.} 2007; Ouriemi, {\it et al.} 2007; Hong, Tao \& Kudrolli 2015; Clark, {\it et al.} 2015).

A spherical particle resting on a rough surface in a linear sheared fluid flow is an important model to understand the threshold of motion of a particle exposed to a fluid flow. { A recent analysis of this model by Lee \& Balachandar (2012) appears to suggest that torques, and not just the forces, can be important to determining the onset of particle motion in sheared fluid flow.} 
In the case of sufficiently low Reynolds numbers, the net hydrodynamic force and torque acting on a sphere attached to a smooth wall in a linear shear flow has been analytically calculated by O'Neill (1968) and Leighton \& Acrivos (1985). At moderate and higher Reynolds number, where fluid inertia is important, analytical results do not exist. However, numerical results have been obtained by Zeng, {\it et al.} (2009) for the drag and lift coefficients acting on a particle attached to a surface at moderate Reynolds numbers. These results indicate that both the lift and moment coefficients about the particle center decrease relative to the drag coefficient with increasing Reynolds numbers. Thus, there is a need for further investigations to quantitatively test { the relative contribution of forces and torques acting on a particle as a function of Reynolds numbers in order to clarify the conditions} needed to dislodge particles in sheared fluid flows.   

Here, we discuss a new experimental system that enables us to quantitatively measure the threshold of motion of a particle in a linearly sheared fluid where the fluid flow and particle motion are visualized to understand its characteristics. Model asperities with well defined pivot points are used to investigate its importance in determining the onset of motion. We demonstrate that the torque balance condition is important to determining the threshold in laminar as well as turbulent sheared flows. Further, we quantitatively describe the critical shear rate required to dislodge a particle as a function of its physical properties using an interpolation of hydrodynamic drag and lift coefficients obtained in the viscous and inertial limits.  

\section{Experimental system}
\begin{figure}
\begin{center}
\includegraphics[width=0.7\linewidth]{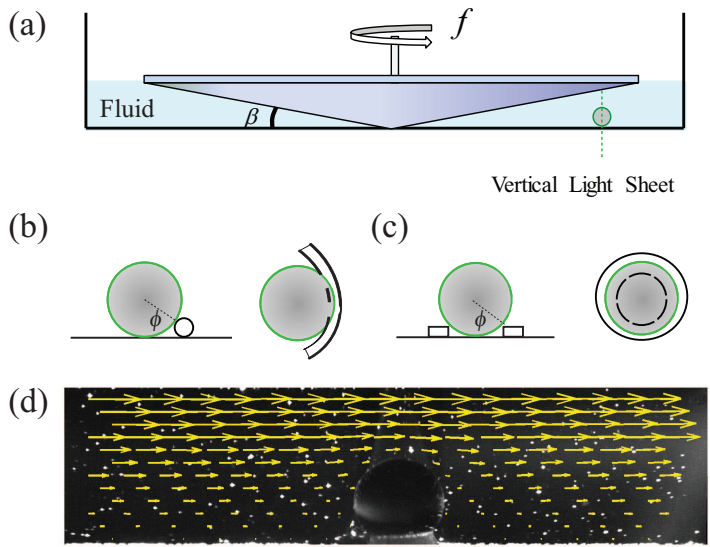}
\end{center}
\caption{\label{fig:model} (a) A schematic diagram of the apparatus. The conical top plate spins about its vertical central axis with a prescribed frequency $f$ resulting in a uniform shear rate $\dot{\gamma} = 2\pi f/\tan\beta$. A vertical light sheet through the particle is used for visualization.  (b) Cross-sectional and top view of a spherical particle lodged against a rod which is bent into a U-shaped pocket. { The line joining the center of the sphere and the point of contact with the asperity makes an angle $\phi$ with respect to the vertical axis.} (c)  Cross-sectional and top view of a spherical particle lodged inside a circular pocket. (d) Flow field observed using Particle Image Velocimetry (PIV) corresponding to $Re_p = 0.1$ superimposed on a sample image. 
}
\end{figure}

A schematic of the experimental apparatus is shown in Fig.~\ref{fig:model}(a). It consists of a stationary transparent cylindrical container with a flat bottom filled with a fluid with a dynamic viscosity $\mu$ and density $\rho_f$,  prepared using water and glycerol mixture ratios reported by Cheng (2008).   Because glycerol-water mixtures are sensitive to temperature, we performed all experiments in a room controlled to within 0.5$^o$C and the viscosity variation within $\pm 2$\%. { This estimate is based on the errors due to the variation in the room temperature and the measurement of fluid volumes used to prepare the glycerol and water mixture. In test experiments, we did not observe any systematic errors in the onsets to within the fluctuations noted for over 3 hours after the fluids were mixed due to  evaporation or hygroscopy.  Therefore, the experiments were all performed well within this time after the fluids were prepared to avoid any such effects.} An inverted cone-shaped top plate, with an apex which coincides with the bottom of the container and radius $R = 95$\,mm, is rotated about its axis with a prescribed frequency $f$ similar to a conical rheometer. { The flows in this geometry are well studied (see for example Sdougos, {\it et al.}, 1984) with the primary flow being concentric with the axis of rotation and increasing linearly from the bottom to the top plate. The  corresponding shear rate is given by $\dot{\gamma} = 2 \pi f/\tan{\beta}$, where $\beta$ is the angle complementary to the cone apex angle. A weak radial secondary flow also occurs which is inward near the flat surface and outward near the cone surface that increases with flow Reynolds number given by $Re = \rho_f \dot{\gamma}R^2 \tan\beta/ 2\mu$, which corresponds to the fluid velocity at the midpoint between the top plate and the bed surface.}

In order to test the relative effects of gravity, inertia and viscosity, we use spherical particles composed of Delrin, polytetrafluoroethylene (PTFE), glass, aluminum, ceramic, titanium, and stainless steel with density $\rho_s = 1400$, 2170, 2500, 2700, 3875, 4512, and 7960 kg\,m$^{-3}$, respectively. While a range of particle diameters were probed, we discuss the data for $d = 3.175 \pm 0.005$\,mm for simplicity of presentation. The Reynolds number at the particle scale $Re_p$ is defined by using the velocity $v = \dot{\gamma}d/2 $ corresponding to the center of the particle. Then,  
\begin{equation}
Re_p = \frac{ \rho_f \dot{\gamma} d^2}{2 \mu } = \frac{ \rho_f \pi f d^2}{\mu \tan{\beta}}. 
\end{equation}

Two kinds of model asperities were used including a rod bent into a U shaped pocket illustrated in Fig.~\ref{fig:model}(b) and a circular pocket using a flat ring illustrated in Fig.~\ref{fig:model}(c). The angle $\phi$ subtended by the line joining the particle center and the pivot point from the vertical characterizes the barrier size relative to the particle size in both cases. The U shape allows the particle to be fully exposed to the fluid flow while also allowing it to move freely in the pocket. This leads the particle to rattle inside the pocket when the flow becomes time-dependent at higher $Re_p$ as shown in the Supplementary Documentation.  In contrast, the circular pocket geometry leads the particle to be confined in all directions inside the pocket and is observed to be stationary at both high and low $Re_p$ before getting dislodged. Because the bottom of the sphere is shielded by the ring, the particle is partially exposed to the fluid flow as in a granular bed.  This leads to an approximate 10\% lowering of $C_d$ and a 7\% lowering of $C_o$ for a circular pocket with $\phi = 38^o$ compared to a fully exposed particle using estimates obtained by Pozrikidis (1997) for low $Re_p$ flows.

\begin{figure}
\begin{center}
\includegraphics[width=0.5\linewidth]{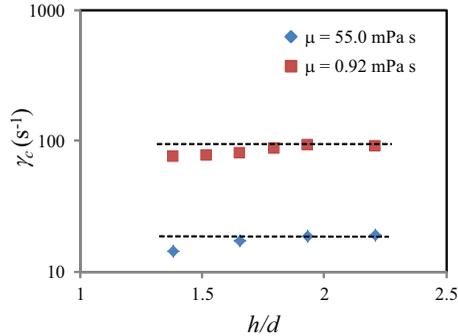}
\caption{\label{fig:gap} The measured critical shear rate $\dot{\gamma_c}$ as a function of fluid height $h/d$ required to dislodge a PTFE particle. We observe that $\dot{\gamma_c}$ is essentially constant for $h/d \gtrsim 2$ in the case of both low and high viscosity fluids used in the experiments. The horizontal dashed lines are guides to the eye. 
}
\end{center}
\end{figure}
Further, because the fluid velocity has to match the velocity of the spinning top boundary, the proximity of the top boundary can lead to a greater drag coefficient at least at low Reynolds numbers compared to unbounded flows (see Happel \& Brenner 1973). We varied the distance $r_b$ { between the particle and} the axis of rotation to understand the effect of the top boundary for a high and low viscosity fluid used in our experiments. Because of the slope of the top surface, this results in a gap height $h = r_b \tan\beta$ between the top boundary and the container bottom. The measured $\dot{\gamma_c}$ as a function of $h$ is shown in Fig.~\ref{fig:gap}. Indeed, we observe that $\dot{\gamma_c}$ is lower for $z/d < 2$, but remains essentially constant for $z/d \gtrsim 2$ for both low and high viscosity fluids used in the experiments. Accordingly, we have confined our discussion to $h = 2d$, which corresponds to the particle being placed at a distance $r_b = 7$\,cm from the axis of rotation, where the direct effect of the top surface can be expected to be small. 

\section{Flow visualization}
\begin{figure}
\begin{center}
\includegraphics[width=0.7\linewidth]{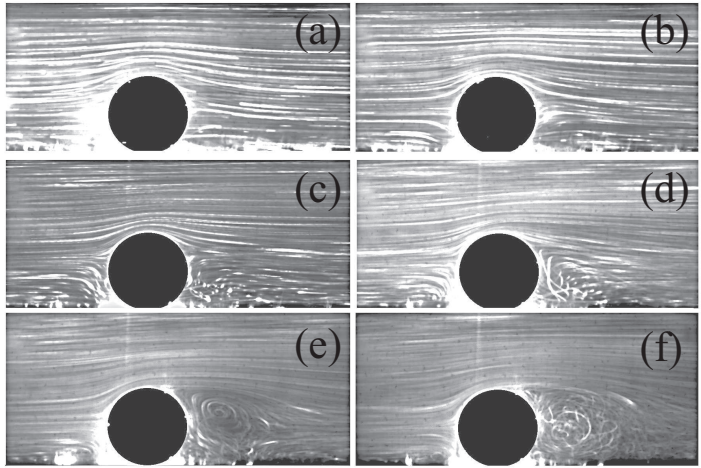}
\end{center}
\caption{\label{fig:piv} The fluid flow observed around a spherical particle glued to the substrate visualized with tracer particles. The exposure time is 1.0\,s. (a) $f = 0.01$\,Hz, $Re = 30$, and $Re_p = 0.7$, (b) $f = 0.03$\,Hz, $Re = 91$, and $Re_p = 2.2$, (c) $f = 0.1$\,Hz, $Re = 304$, and $Re_p = 7.3$, and (d) $f = 0.3$\,Hz, $Re = 916$, and $Re_p = 22$, (e) $f = 1.0$\,Hz, $Re = 3041$,  and $Re_p = 73$, and (f) $f = 3.0$\,Hz, $Re = 9160$, and $Re_p = 220$.  The flow is  symmetric at low $Re_p$ and vortices develop as $Re_p$ is increased and a vortex clearly develops in the wake at $Re_p \sim 20$. The flow in the wake clearly becomes time dependent at $Re_p = 220$. A smaller vortex in front of the sphere is observed for $Re_p \ge 22$. (See movies in Supplementary Documentation.) }
\end{figure}

In order to characterize the nature of the flow, we performed experiments with  tracer particles added to a water-glycerol mixture corresponding to $\rho_f = 1100$ kg m$^{-3}$ and $\mu = 3$\,mPa\,s in which the tracers are neutrally buoyant. The fluid was illuminated with a light sheet which transects the particle in a vertical plane through its center. { In case of sufficently slow flows, we use Particle Image Velocimetry (PIV) to obtain the fluid flow with a sequence of images acquired at 2 frames per second, and analyzing the images using the shareware computer program OpenPIV (http://www.openpiv.net/). 
An example of which is shown in Fig.~\ref{fig:model}(d) for low $Re_p$. 
The fluid velocity shown by the length of the arrows can be noted to be symmetric about the vertical axis and increase linearly in regions away from the sphere. The measured shear rate from PIV is found to be within 5.5\% of that calculated using the rotation rate of the top plate}. 

The flow structure at higher $Re_p$ can be deduced by examining the streaks made by the tracers over one second in Fig.~\ref{fig:piv}  and the corresponding movies in the Supplementary Documentation. One can observe that the flow is essentially symmetric for $Re_p \sim 1$, but grows asymmetric as $Re_p$ is increased.  A vortex can be clearly observed at $Re_p \sim 20$ and the wake grows and becomes time dependent for  $Re_p \sim 220$. Further, one also observes the development of a smaller vortex for $Re_p \ge 22$ in the front of the sphere near the substrate (see Movie in Supplementary Documentation). In all cases, the flow well in front of the particle appears to be laminar, and the eddies generated by the flow around the particle have decayed by the time the fluid flow returns after going around the circular track over the range of $Re_p$ visualized.

 We also tested the effect of the secondary flows that can arise in this system at high $Re$. In particular, we  examined the departure angle of the particles from the azimuthal direction when they are dislodged over a U-shaped barrier by visualizing the system through the transparent bottom of the container. By measuring the departure angle for all the particles used in our study, we find angles from the azimutal direction to $3 \pm 4.5^o$ when $Re_p$ is varied between 40 to 243. Such a variation would lead to less than 2\% underestimation of the shear rate at onset which is within the experimental error. Therefore, we conclude that secondary flows are negligible in determining the main trends observed in the study. 

\section{Measured critical shear rate}
\begin{figure}
\begin{center}
\includegraphics[width=0.7\linewidth]{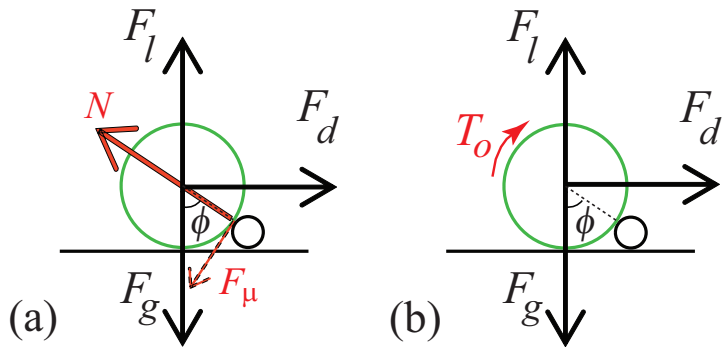}
\end{center}
\caption{\label{fig:dia} {(a)  A schematic diagram representing the forces acting on the particle resting against an asperity with a circular crosssection. Contact forces can be considered to be absent at the point of contact between the particle and the substrate at onset. (b) A schematic diagram representing the forces as well as the torque $T_o$ that act about the center of the particle. The contact forces acting at the pivot point are not drawn for clarity (see text).} }
\end{figure}

\begin{figure}
\begin{center}
\includegraphics[width=0.55\linewidth]{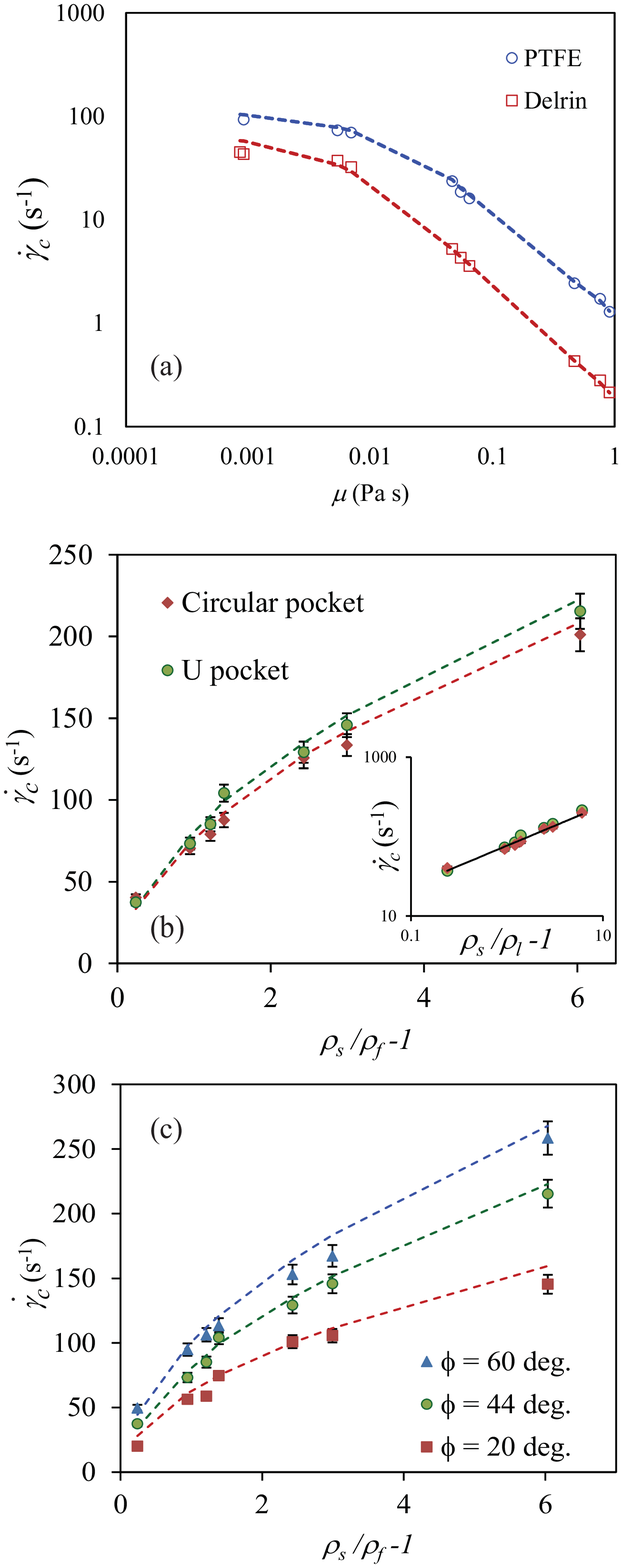}
\end{center}
\caption{\label{fig:rawdata} (a) The critical shear rate $\dot{\gamma_c}$ is observed to decrease with viscosity ($\phi = 44^o$). The fluid viscosity is obtained by using water glycerol mixture ratios reported by Cheng (2008). The error bars are of the order of the symbol size.  (b) $\dot{\gamma_c}$ increases nonlinearly, irrespective of the shape of the asperity. Inset: Same plot in log-log scale with a slope 1/2 line to guide the eye. $\dot{\gamma_c}$ increases consistent with $\sqrt{\rho_s/\rho_f -1}$ for both kinds of barriers. (c) $\dot{\gamma_c}$ is systematically greater for higher $\phi$. The dashed lines in the plots correspond to $\dot{\gamma_c}$ calculated using Eq.~\ref{eq:freq} and  Eq.~\ref{c_h} with $\alpha_0 = 0.45$ and $\alpha_d = 0.65$. } 
\end{figure}

With this characterization of the flow, we now discuss the measured critical shear rate $\dot{\gamma_c}$ as a function of experimental control parameters. 
Figure~\ref{fig:rawdata}(a) shows $\dot{\gamma_c}$ as a function of $\mu$ corresponding to a U-shaped pocket with $\phi = 44^o$. Each data point corresponds to three independent measurements and the error is less than 5\%.  The data was obtained by linearly increasing the rotation frequency of the top plate to a prescribed value $f$ and holding it constant for a fixed wait time of 100\,s. The threshold is reached if the particle is observed to roll out over the barrier during this wait time interval.  The particle was observed to move and dislodge immediately after threshold was reached at low $Re_p$. However,  the particle was observed to rattle inside the pocket and dislodge after a few seconds in the case of the U-shaped pocket when $Re_p \gtrsim 10$. In the case of a circular pocket, no such rattling was observed and the particle dislodged right after first moving. We also found that using a longer wait time did not lead to a systematic change in the measured threshold. But, decreasing the wait time increased the threshold somewhat. Because we are interested in the long time behavior, we have used a wait time of 100\,s for consistency. 

We observe that $\dot{\gamma_c}$ decreases systematically because the drag can be expected to increase with $\mu$. Further, $\dot{\gamma_c}$ is observed to be systematically higher for PTFE because it has a higher density compared with Delrin. To further probe the trends with respect to the density of the particle, we plot $\dot{\gamma_c}$ as a function of $\rho_s/\rho_f -1$ in Fig.~\ref{fig:rawdata}(b) corresponding to $\mu = 5.2$\,mPa\,s, and where the data corresponds to $Re_p > 10$. From the log-log plot in the corresponding inset, we observe that $\dot{\gamma_c}$ increases consistent with a square root function. We have also plotted data corresponding to a barrier with a circular pocket, and we observe similar trends.  Finally, we have plotted $\dot{\gamma_c}$ in Fig.~\ref{fig:rawdata}(c) for various $\phi$ and find higher $\dot{\gamma_c}$  for higher $\phi$. Thus, $\dot{\gamma_c}$ increases  systematically with greater barrier height. 

\section{Conditions to dislodge particle}
To explain these observations, we next discuss the gravitational and hydrodynamic forces and torques acting on the particle which are used to determine the condition for stability. The gravitational force acting on the particle including the effect of buoyancy is given by
$F_g = \frac{1}{6} \pi (\rho_s - \rho_f) g d^3$,  
where $g$ is the acceleration due to gravity, and the corresponding torque due to gravity about the pivot point on the barrier is given by 
$T_g = F_g \frac{d}{2} \sin{\phi}.$ The net drag force acting on a sphere can be written as
$F_d = \frac{1}{8} C_d \rho_f v^2 \pi d^2$, 
where $C_d$ is the drag coefficient which depends on $Re_p$. 
The torque due to drag about the center of the particle can be written as
$T_o = \frac{1}{16} C_o \rho_f v^2 \pi d^3$,
where $C_o$ is a drag coefficient which also depends on $Re_p$.
Then, the torque due to drag $T_d$ about the pivot point can be written as the sum of the torque about the center and the net force times the projected distance from the center to the pivot point, i.e. $T_d = T_o + F_d \frac{d}{2} \cos{\phi}$. The lift due to the difference of flow velocity above and below the particle center, can be written as 
$F_l = \frac{1}{8} C_l \rho_f v^2 \pi d^2$, where $C_l$ is the lift coefficient, and the corresponding torque due to the lift $T_l = F_l \frac{d}{2} \sin{\phi}$. 
{ Because of the fore/aft asymmetry that develops in the flow as shown in Fig.~\ref{fig:piv}, one can further expect the effective point where the lift and drag act to shift from the vertical axis of symmetry.  However,  we
are unaware of any previous work which discusses this effect, and we assume that the lift acts at the center for simplicity.}{ Further, the particle also experiences normal reaction force and tangential friction forces at contact points with the substrate and the barrier. One may expect these contact forces to approach zero at the point of contact between the particle and the bottom substrate, just when the particle is about to be dislodged. However, reaction force $N$ and friction force $F_\mu$ can be expected to be present at the point of contact between the particle and the barrier even as the particle is dislodged.} 

\subsection{ Sliding Contact}
{ We first consider the force components perpendicular to the line joining the particle center and the contact point between the particle and the barrier as shown in Fig.~\ref{fig:dia}(a) assuming that the particle is dislodged by sliding over the barrier in the direction opposite to $F_\mu$. Then, the  force equation for equlibrium  
gives}  $F_d \cos{\phi} + F_l \sin{\phi} - F_\mu = F_g \sin{\phi}$, where, $F_\mu$ is the friction force at the point of contact.  By rearranging terms, one sees that 
$(F_d \cos{\phi} + F_l \sin{\phi})/ F_g \sin{\phi} = 1 + F_\mu/F_g \sin{\phi}.$ 
Because of the presence of the fluid, and because it is difficult to determine the degree to which the particle rolls versus slides at the point of contact, $F_\mu$ is difficult to estimate with certainty. { Nonetheless, the second term on the right hand side can be assumed to be positive if not zero in the case where $F_\mu$ goes to zero. Thus,}
\begin{equation}
(F_d \cos{\phi} + F_l \sin{\phi})/ F_g \sin{\phi} \geq 1. 
\label{eq:forcec}
\end{equation} 

\subsection{ Rolling Contact}
{ However, from the movies shown in the supplemetary documentation, one notes that the particle appears to roll rather than slide when it is dislodged by the flow. } Therefore, it appears that the sliding friction leads the particle to pivot about the point of contact. Then, the condition for torque balance about the pivot point is given by 
\begin{equation}
T_d + T_l = T_g, 
\label{eq:torque}
\end{equation}
where, both $T_d$ and $T_l$ can be seen to act clockwise in the inset to Fig.~\ref{fig:dia}(b) to dislodge the particle, while $T_g$ acts in the counterclockwise direction and keeps the particle from being dislodged. { Because the contact forces act at the pivot point, they do not appear in Eq.~\ref{eq:torque}.} We substitute $T_d$,  $T_l$, and $T_g$ with their expressions in terms of $C_d$, $C_o$, $C_l$, $\rho_s$ and $\rho_f$  in Eq.~\ref{eq:torque} for torque balance. Then, introducing a net hydrodynamic coefficient $C_h = C_o + C_d \cos\phi + C_l \sin\phi$ and recalling that $v = \dot{\gamma_c}d/2$, we obtain the shear rate required to dislodge a particle as  
\begin{equation}
\dot{\gamma_c} = \sqrt{\frac{16 (\rho_s - \rho_f) g \sin{\phi}}{3 \rho_f C_h d} }. 
\label{eq:freq}
\end{equation}
Thus, for a given particle, fluid and surface roughness, $\dot{\gamma_c}$ can be evaluated provided $C_h$ is known for that $Re_p$. Alternately, $C_h$ can be determined by rewriting Eq.~\ref{eq:freq} as 
\begin{equation}
C_h = \frac{16 (\rho_s - \rho_f) g \sin \phi}{3 \rho_f \dot{\gamma_c^2} d},
\label{eq:c_h}
\end{equation}
where, all the quantities on the right hand side can be measured in our experiments.  

\begin{figure}
\begin{center}
\includegraphics[width=0.7\linewidth]{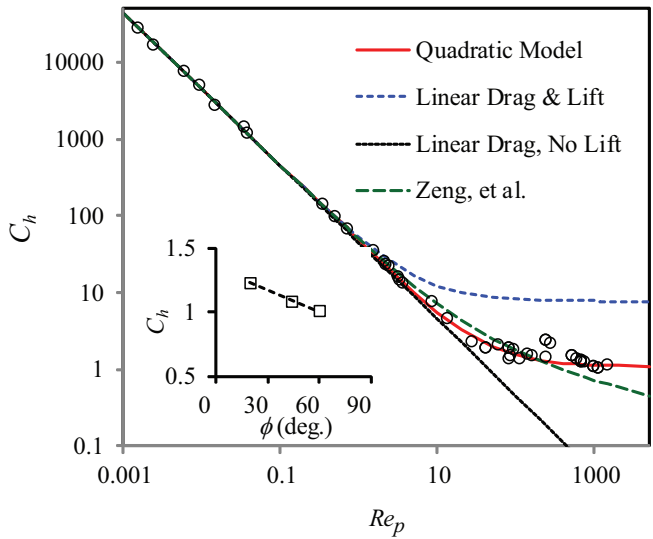}
\end{center}
\caption{\label{fig:Ch} The net hydrodynamic coefficient $C_h$ obtained from the experiment as a function of $Re_p$ for $\phi =  43.5^o$ compared with various models. Here, the curve labeled as the quadratic model includes the quadratic drag contribution in calculating $C_h$ (see Eq.~\ref{c_h}). The curves corresponding to linear drag use O'Neill's form for $C_d$, and the lift is obtained using Leighton and Acrivos's form for $C_l$ (see text).  $C_h$ is observed to approach a constant at the highest $Re_p$  ($\alpha_0 = 0.45$ and $\alpha_d = 0.65$). { The coefficients used to generate the curve by Zheng, {\it et al.} (2009) are predicted up to $Re_p \sim 200$. However, systematic deviations are observed above $Re_p \sim 10$.}  Inset: $C_h$ decrease somewhat for higher $\phi$ ($Re_p \sim 1000$). The line corresponds to Eq.~\ref{eq:c_h} with $Re_p =1000$. 
}
\end{figure}

The measured $C_h$ using Eq.~\ref{eq:c_h} is plotted in Fig.~\ref{fig:Ch} as a function of $Re_p$ also over a wide combination of particle densities and viscosities for a fixed roughness. We find that the measured $C_h$ decreases linearly for low $Re_p$ before rapidly approaching a constant value at the highest $Re_p$ studied. We have further plotted $C_h$ for $Re_p \sim 1000$ for the three different $\phi$ studied in the inset to Fig.~\ref{fig:Ch}. It can be observed that $C_h$ becomes relatively independent of $Re_p$, while systematically decreasing with $\phi$. 

In the viscous limit, the total drag force and torque acting on a particle attached to a wall in a linear shear flow has been calculated by O'Neill (1968). Assuming drag to be linear with velocity, he found $C_d$ and $C_o$ to be 
$C_d^0 = \frac{24 f_w}{Re_p}$, 
and    
$C_o^0 = \frac{16 b_w}{Re_p}$,
where, $f_w = 1.7005$ and $b_w = 0.944$ are constants which arise due to the no-slip boundary condition at the substrate. 
Further, a lift coefficient $C_l^0 = 6.888 f_w$ has been calculated corresponding to viscous shear lift for low $Re_p$ by Leighton and Acrivos (1985). We compare $C_h$ using these hydrodynamic coefficients in Fig.~\ref{fig:Ch} with those obtained directly from our experimental measurements. We find good agreement for $Re_p < 1$ whether or not we include lift in the calculations. 

Extrapolating the curves into the inertial regime, the two curves deviate systematically above or below the data, depending on whether we consider lift or not. 
In fact, Zeng, {\it et al.} (2009) have found $C_d = \frac{24 f_w}{Re_p} (1 + 0.104 Re_p^{0.753})$ based on numerical simulations for a fixed sphere on an infinite plane which is linearly sheared by a fluid flow when $Re_p \leq 250$. They also found that $C_o^0$ can be extended up to $Re_p = 200$. Further, they postulated that the lift coefficient can be interpolated between the low and high $Re_p$ limits as $C_l = {3.663}{(Re_p^2 + 0.1173)^{-0.22}}$  for $Re_p < 200$, although it may be noted that higher lift has been also measured over the same regime by Mollinger \& Nieuwstadt (1996).  Accordingly, we have calculated $C_h$ and plotted the result in Fig.~\ref{fig:Ch}. The corresponding curve appears to capture the overall trend in the data to $Re_p \sim 10$ within experimental error.  {  However, systematic deviations can be observed above this value over the range of validity of those simulations.}

\begin{figure}
\begin{center}
\includegraphics[width=0.5\linewidth]{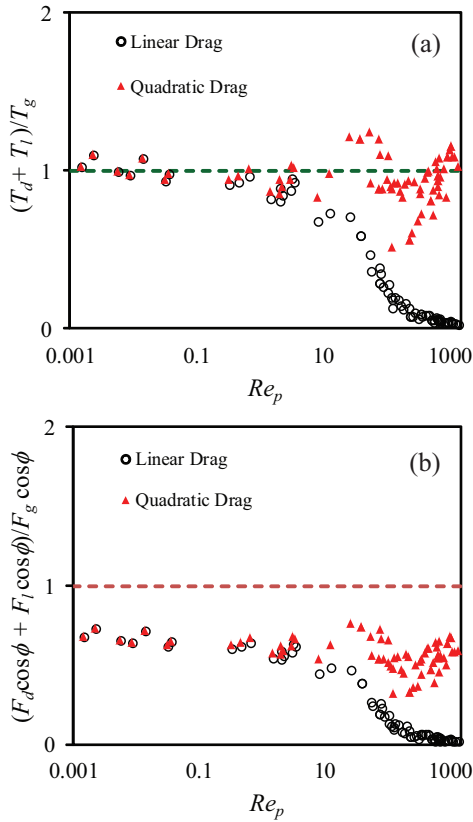}
\end{center}
\caption{\label{fig:high_re} (a) The ratio of the torques associated with hydrodynamic forces and gravity calculated from the measured $\dot{\gamma_c}$ and using the linear drag model, and the quadratic drag model in Eq.~\ref{c_h}.  The horizontal dashed line corresponds to the torque balance condition given by Eq.~\ref{eq:torque}. Good agreement is observed with analytical calculations in the $Re_p < 1$ regime. Overall good agreement is also observed with the torque balance condition by using a quadratic model over the entire range of $Re_p$. (b) The ratio of the force components given by the left hand side of Eq.~\ref{eq:forcec} plotted versus $Re_p$. { The drag and lift are obtained using the measured shear rate required to dislodge the particle. The measured ratio is clearly below the horizontal dashed line showing that the torque about the particle needs to be taken into account to describe onset of motion.} } 
\end{figure}

In order to describe the data over the entire $Re_p$ measured,  we consider $C_d$ and $C_o$ as a superposition of analytically calculated coefficients in the low $Re_p$ limit and a term corresponding to quadratic drag which is independent of $Re_p$, i.e.     
$C_d = C_d^0 + \alpha_d$,
and,
$C_o = C_o^0 + \alpha_o$,
where, $\alpha_d$ and $\alpha_o$ depend further on the flow geometry. 
The lift acting on a particle attached to a wall at high $Re_p$ has been measured to be $0.242$ (see Okamoto 1979) which we round up to be $\alpha_l = 0.25$. We then interpolate $C_l$ between the viscous shear lift at low $Re_p$ and the lift at high $Re_p$ using the function $C_l  = (C_l^0 - \alpha_l) \exp{(-Re_p)} + \alpha_l$.  { Accordingly, we postulate that 
\begin{equation}
C_h = \big[\frac{16 b_w}{Re_p} + \alpha_o\big] + \big[\frac{24 f_w}{Re_p}  + \alpha_d\big] \cos\phi + \big[(6.888 f_w - \alpha_l) \exp{(-Re_p)} + \alpha_l\big]\sin\phi\,,
\label{c_h}
\end{equation}
where, the first, second, and third term in brackets on the right hand side correspond to the interpolated moment, drag, and lift acting relative to the particle center.  Then, we obtain $\alpha_o = 0.5 \pm 0.1$ and $\alpha_d = 0.7 \pm 0.1$ by fitting Eq.~\ref{eq:c_h} to the data shown in the inset to Fig.~\ref{fig:Ch}. The fitted value of $\alpha_d$ is much greater than for a sphere in uniform unbounded flow, but consistent with measurements of drag coefficent of $C_d \approx 0.627$ reported for a sphere attached to a surface at high $Re_p$ obtained by measuring the surface-pressure distribution on a sphere by Okamoto (1979).} It may be noted that $C_h$ is not very sensitive to $\phi$ at high $Re_p$. This occurs because the decrease in drag contribution to torque is compensated by an increase in the lift contribution  as $\phi$ is increased.

We have plotted $\dot{\gamma_c}$ obtained using Eq.~\ref{eq:freq} and Eq.~\ref{c_h} in all three plots in Fig.~\ref{fig:rawdata}. We find that by including the quadratic drag in the inertial regime and combining it with the analytical results  in the viscous regime, we are able to capture the observed $\dot{\gamma_c}$ dependence as a function of fluid viscosity, particle density, and surface roughness characterized by $\phi$. { We further tested to see if assuming the forces alone can describe the data.} This corresponds to forcing $C_o$ to be zero in our fits, and we are simply not able to obtain an accurate description of these trends.   

In order to summarize the results, we have plotted the ratio of the torques due to hydrodynamic forces and gravity in Fig.~\ref{fig:high_re}(a) and the ratio of the hydrodynamic forces and gravity in Fig.~\ref{fig:high_re}(b) as a function of $Re_p$. We find that the threshold condition is well described by the analytically calculated torques in the viscous limit for $Re_p < 0.5$ before inertial effects grow, leading to systematic deviations. The form of drag given by the viscosity and inertia components captures the data relatively well over the entire range of $Re_p$ investigated. By contrast, the data  shown in Fig.~\ref{fig:high_re}(b) falls systematically below one, which is the lower bound given by Eq.~\ref{eq:forcec} even if one assumes $F_\mu =0$. Thus, the critical shear rate remains distinct and well below the threshold condition { obtained by considering the forces alone} to dislodge the particle. Based on this observation, we conclude that the condition when the particle gets dislodged clearly corresponds to the torque balance condition. { This is further in agreement with the observation that the particle rolls over the barrier when the corresponding critical shear rate is reached as illustrated by the movies in the Supplementary Documentation.} 

{ In performing this analysis, we have assumed that the shear rate  required to dislodge the particle is given by the critical rotation frequency of the top plate.  It is possible that this method can lead to a systematic overestimation of the hydrodynamic coefficients when the flow becomes time-dependent at higher $Re_p$. However, this systematic error is offset by the corresponding lower estimate of shear rates used in the calculation of torques and forces. Thus, we expect the hydrodynamic forces and torques used in Fig.~\ref{fig:high_re} to be robust even at high $Re_p$.} 

\section{Conclusions}

In summary, we have shown with experiments that the shear rate at onset of erosion is determined by the torque balance condition. { Further, systematic deviations are observed if forces alone are considered in determining the instability of the particle.} The main reason for this discrepancy is because the net hydrodynamic force does not act at the center of the particle but rather some distance above the particle center because of the fact that the mean flow speed increases with distance from the bottom substrate. In the torque balance condition, this is taken into account by considering the additional torque about the center of the particle. Building on this condition, we have then quantitatively described the observed critical shear rate $\dot{\gamma_c}$ on the particle density, the fluid viscosity, and the surface roughness over a wide range of particle Reynolds Numbers $Re_p$.  We find that a linear combination of the hydrodynamic coefficients obtained in the viscous and inertial limits can describe the observed $\dot{\gamma_c}$ as a function of the particle and fluid properties from laminar to turbulent flow conditions.

Further, we show that the data at low $Re_p < 0.5$ is in good agreement with analytical calculations of the drag and lift coefficients in the $Re_p \rightarrow 0$ limit, but differ from numerical results at moderate $Re_p$ reported by Zeng, {\it et al.} (2009) for flow past a sphere resting on a smooth surface.  At higher $Re_p$, where analytical results are unavailable, the hydrodynamic coefficients are found to approach a constant for $Re_p > 1000$.  It is possible that the differences from the numerical results at moderate $Re_p$ arise because of the presence of the physical barriers near the base of the particle in the experiments which can modify the flow.  Further research is required to fully understand the effect of surface roughness and particle exposure to extend the implications of our study to the erosion of a granular bed as in rivers and streams.    

\section{Acknowledgments}

This material is based upon work supported by the U.S. Department of Energy Office of Science and Office of Basic Energy Sciences program under DE-SC0010274, and National Science Foundation Grant  No. CBET-1335928.

\bibliographystyle{jfm}

\begin{thebibliography}{23}

\bibitem{Buffington} Buffington, J. M.   and Montgomery, D. R.  1997 ``A systematic analysis of eight decades of incipient motion studies, with special reference to gravel bedded rivers," Water Resources Research {\bf 33}, 1993.

\bibitem{Charru} Charru, F. Mouilleron,  H. and Eiff, O.  2004 ``Erosion and deposition of particles on a bed sheared by a viscous flow," Journal of Fluid Mechanics {\bf 519}, 55.

{ \bibitem{Charru07} Charru, F. Larrieu, E. Dupont, J.-B and Zenit, R.  2007 ``Motion of a particle near a rough wall in a
viscous shear flow," Journal of Fluid Mechanics {\bf 570}, 431.}

\bibitem{cheng08} Cheng, N.-S.  2008 ``Formula for the Viscosity of a Glycerol-Water Mixture," Industial \& Engineering Chemistry Research {\bf 47}, 3285. 

\bibitem{clark15} Clark, A. H. Shattuck, M. D. Ouellette, N. T. and O'Hern, C. S.  2015 ``Onset and cessation of motion in hydrodynamically sheared granular beds," Physical Review E {\bf 92}, 042202. 


\bibitem{happel73} Happel, J. and Brenner, H. 1973 {\it Low Reynolds Number Hydrodynamics}, 2nd edn. Noordhoff, Leyden. 

\bibitem{hong2015} Hong, A. Tao,  M. J. and Kudrolli, A.  2015  ``Onset of erosion of a granular bed in a channel driven by fluid flow," Physics of Fluids {\bf 27}, 013301.

\bibitem{lee12} Lee, H. and Balachandar, S. 2012 ``Critical shear stress for incipient motion of a particle on a rough bed," Journal of Geophysical Research: Earth Surface {\bf 117}, F1. 

\bibitem{Leighton} Leighton, D.  and Acrivos, A.  1985 ``The lift on a small sphere touching a plane in the presence of a simple shear flow," Zeitschrift f\"ur angewandte Mathematik und Physik ZAMP {\bf 36}, 174. 

\bibitem{Mollinger} Mollinger, A. M.  and Nieuwstadt, F. T. M.  1996  ``Measurement of the lift force on a particle fixed to the wall in the viscous sublayer of a fully developed turbulent boundary layer," Journal of Fluid Mechanics {\bf 316}, 285.

\bibitem{O'neill} O'Neill, M. E. 1968 ``A sphere in contact with a plane wall in a slow linear shear flow," Chemical Engineering Science {\bf 23}, 1293. 

\bibitem{okamoto79} Okamoto, S. 1980 ``Turbulent Shear Flow Behind a Sphere Placed on a Plane Boundary," in: Turbulent Shear Flows 2, ed. by L.J.S. Bradbury, F. Durst, B.E. Launder, F.W. Schmidt, J.H. Whitelaw (Springer, Berlin Heidelberg New York) 246.

\bibitem{Ouriemi} Ouriemi, M. Aussillous,  P. Medale, M.  Peysson, Y.  and Guazzelli, E.  2007 ``Determination of the critical Shields number for particle erosion in laminar flow," Physics of Fluids {\bf 19}, 061706.

\bibitem{phillps80} Phillips, M.  1980 ``A force balance model for particle entrainment into a fluid stream," Journal of Physics D: Applied Physics {\bf 13}, 221.

\bibitem{pozrikidis97} Pozrikidis, C.  1997 ``Shear flow over a protuberance on a plane wall," Journal of Engineering Mathematics {\bf 31}, 29.


\bibitem{Shields} Shields, A.  1936 ``Anwendung der \"Ahnlichkeitsmechanik und der Turbulenz-forschung auf die Geschiebebewegung," Preussische Versuchsanstalt f\"ur Wasserbau und Schiffbau {\bf 26}, 524.

{ \bibitem{sdougos1984} Sdougos, H.P. Bussolari,   S.R. and Dewey, C.F. 1984  ``Secondary flow and turbulence in a cone-and-plate device," Journal of Fluid Mechanics {\bf 138}, 379. }

\bibitem{wiberg87} Wiberg, P. L. and Smith, J. D.  1987 ``Calculations of the critical shear stress for motion of uniform and heterogeneous sediments," Water Resources Research {\bf 23}, 1471.

\bibitem{Zeng} Zeng, L. Najjar,  F. Balachandar, S. and Fischer, P. 2009 ``Forces on a finite-sized particle located close to a wall in a linear shear flow," Physics of Fluids {\bf 21}, 033302. 



\end{thebibliography}

\end{document}